\documentclass{aa}  
\usepackage{graphicx}	
\usepackage{amsmath}	
\usepackage{amssymb}	
\usepackage{booktabs}
\usepackage{multirow}

\usepackage{lineno}
\usepackage{subcaption}
\usepackage{float}
\usepackage{txfonts}
\usepackage{xcolor}

\definecolor{myblue}{rgb}{0.00, 0.0, 0.9}
\definecolor{myred}{rgb}{0.90, 0.0, 0.0}
\definecolor{mygreen}{rgb}{0.0, 0.7, 0.0}
\usepackage[breaklinks,  citecolor=myblue, linkcolor=myred, urlcolor=purple, colorlinks=true, debug, baseurl=' ']{hyperref}
\usepackage{cleveref}

\begin{document}

   \title{Bright-Moon Sky as a Wide-Field Linear Polarimetric Flat Source for Calibration}


   \author{S.\,Maharana\inst{1,2,3}
          \and
          S.\,Kiehlmann\inst{1,2}
          \and
          D.\,Blinov\inst{1,2}
          \and
          V.\,Pelgrims\inst{1,2}
          \and
          V.\,Pavlidou\inst{1,2}
          \and
          K.\,Tassis\inst{1,2}
          \and
          J.\,A.\,Kypriotakis\inst{1,2}
          \and
          A.\,N.\,Ramaprakash\inst{1,3,4}
          \and
          R.\,M.\, Anche\inst{14}
          \and
          A.\,Basyrov\inst{5}
          \and
          K.\,Deka\inst{6,7}
          \and
          H.\,K.\,Eriksen\inst{5}
          \and
          T.\,Ghosh\inst{6}
          \and
          E.\,Gjerløw\inst{5}
          \and
          N.\,Mandarakas\inst{1,2}
          \and
          E.\,Ntormousi\inst{1,2}
          \and
          G.\,V.\,Panopoulou\inst{4,8}
          \and
          A.\,Papadaki\inst{1,2,9,10}
          \and
          T.\,Pearson\inst{4}
          \and
          S.\,B.\,Potter\inst{11,12}
          \and
          A.\,C.\,S.\,Readhead\inst{1,13}
          \and
          R.\,Skalidis\inst{13}
          \and
          I.\,K.\,Wehus\inst{5}
          }

   \institute{Institute of Astrophysics, Foundation for Research and Technology - Hellas, Vasilika Vouton, GR-70013 Heraklion, Greece\\
              \email{sid@ia.forth.gr, sidh345@gmail.com}
         \and
            Department of Physics, University of Crete, Voutes University Campus, GR-70013 Heraklion, Greece
        \and
            Inter-University Centre for Astronomy and Astrophysics, Post Bag 4, Ganeshkhind, Pune - 411 007, India             
        \and
            Cahill Center for Astronomy and Astrophysics, California Institute of Technology, Pasadena, CA, 91125, USA
        \and
            Institute of Theoretical Astrophysics, University of Oslo, P.O. Box 1029 Blindern, NO-0315 Oslo, Norway
        \and
            National Institute of Science Education and Research, An OCC of Homi Bhabha National Institute, Bhubaneswar - 752050, India
        \and
            Astrophysics Division, National Centre for Nuclear Research, Pasteura 7, Warsaw, 02093, Poland
        \and
            Department of Space, Earth \& Environment, Chalmers University of Technology, SE-412 93 Gothenburg, Sweden
        \and
            Institute of Computer Science, Foundation for Research and Technology-Hellas, Vasilika Vouton, GR-70013 Heraklion, Greece
        \and
            Department of Computer Science, University of Crete, Voutes, 70013 Heraklion, Greece
        \and 
            South African Astronomical Observatory, PO Box 9, Observatory, 7935, Cape Town, South Africa
        \and
            Department of Physics, University of Johannesburg, PO Box 524, Auckland Park 2006, South Africa
        \and 
            Owens Valley Radio Observatory, California Institute of Technology, Pasadena, CA, 91125, USA
        \and
            Steward Observatory, University of Arizona, Tucson, Arizona, 85721, USA
             }

   \date{Received September 15, 1996; accepted March 16, 1997}

 
  \abstract
{Next-generation wide-field optical polarimeters like the Wide-Area Linear Optical Polarimeters (WALOPs) have a field of view (FoV) of tens of arcminutes. For efficient and accurate calibration of these instruments, wide-field polarimetric flat sources will be essential. Currently, no established wide-field polarimetric standard or flat sources exist.}
{This paper tests the feasibility of using the polarized sky patches of the size of around ten-by-ten arcminutes, at a distance of up to $20^{\circ}$ from the Moon, on bright-Moon nights as a wide-field linear polarimetric flat source. }
{We observed 19 patches of the sky adjacent to the bright-Moon with the RoboPol instrument in the SDSS-r broadband filter. These were observed on five nights within two days of the full-Moon across two RoboPol observing seasons.}
{We find that for 18 of the 19 patches, the uniformity in the measured normalized Stokes parameters $q$ and $u$ is within 0.2~\%, with 12 patches exhibiting uniformity within 0.07~\% or better for both $q$ and $u$ simultaneously, making them reliable and stable wide-field linear polarization flats.}
{We demonstrate that the sky on bright-Moon nights is an excellent wide-field linear polarization flat source. Various combinations of the normalized Stokes parameters $q$ and $u$ can be obtained by choosing suitable locations of the sky patch with respect to the Moon}


   \maketitle
%

\section{Introduction}
Optical polarimetry is a powerful diagnostic tool that has been used by astronomers to probe many astrophysical objects, especially systems where there is an asymmetry in light emission and/or propagation. Some commonly studied objects through optical polarimeters include active galactic nuclei, novae and supernovae, and dust clouds in the interstellar medium (ISM) (e.g., \citealt{Hough_review}; \citealt{Scarrott-1991}; \citealt{trippe2014polarization}). Polarimeters are often designed to achieve accuracies of $p \lesssim 0.1$~\% or better with careful calibration observations to estimate the instrument-induced polarization. Most polarimeters built to date are optimized for observation of either point sources or very narrow fields of view (FoV) of a few arcminutes. The calibration of these polarimeters is done using measurements of polarimetric standard stars, as described in papers reporting the commissioning and performance of various past instruments (e.g., \citealt{impol}; \citealt{robopol}; \citealt{HOWPol}; \citealt{salt_commisioning}; \citealt{Tinyanont2018}; \citealt{DIPOL2}; \citealt{MIMIR}).

\par Like many other fields in astronomy, optical polarimetry is entering an era of large sky surveys with programs like \textsc{Pasiphae}  (Polar-Areas Stellar Imaging in Polarization High-Accuracy Experiment, \citealt{tassis2018pasiphae}), SouthPol \citep{SouthPol} and VSTpol \citep{vst_pol} currently under development. All these surveys will be using unprecedentedly large FoV (> 0.25 square degrees) polarimeters as their main workhorse instruments and are aiming to achieve polarimetric accuracy of  $p \lesssim 0.1$~\% to enable the tomographic reconstruction of the dusty magnetized ISM (\citealt{Pelgrims2023}), among other science cases. Of these, \textsc{Pasiphae} will be concurrently carried out from the Northern and Southern hemispheres using two WALOP (Wide-Area Linear Optical Polarimeter) instruments. The first of the two WALOPs, WALOP-South will be mounted on the South African Astronomical Observatory's 1~m telescope at the Sutherland Observatory and is scheduled for commissioning in 2023. \cite{walop_s_spie_2020, WALOP_South_Optical_Design_Paper} provide a detailed description of the optical and optomechanical design of the WALOP-South instrument.

\par The goal of the WALOP-South instrument is to achieve polarimetric measurement accuracy of $p \lesssim 0.1$~\% across a FoV of $35\times35$~ arcminutes. Complete modeling of the instrument's polarization behavior, as well as the development of the on-sky calibration method, has been completed and presented in \cite{WALOP_Calibration_paper} (to be referred to as Paper I from hereon) and \cite{WALOP_Stress_Birefringence}. The calibration model for the WALOP-South instrument uses the following two ingredients: (a) an in-built calibration polarizer at the beginning of the instrument, and (b) multiple on-sky linear polarimetric flat sources of size $10\times10$~arcminutes or more at various polarization angles (Electric Vector Position Angle, $\theta$), i.e., the polarization values are spread across the $q-u$ plane. We refer the reader to Paper I for a detailed description of the calibration method. 

Currently, while there are multiple known polarized and unpolarized standard stars \citep{robopol_data_release}, they are scattered across the sky and, alone, they are unsuitable for wide-field instrument calibration. While unpolarized wide-field regions can be predicted based on ISM extinction \citep{Skalidis}, finding uniformly polarized regions is harder as it requires long-term monitoring of hundreds of stars, unfeasible with currently available limited FoV polarimeters. As mentioned, for WALOP instruments, in particular, multiple wide-field polarized sources, spread over the $q-u$ planes are needed. Standard polarized regions whose polarization values are known a priori would be ideal, but knowledge about the polarization value is not a critical requirement. Rather, linear polarimetric flat regions, which have a constant polarization across the region, are sufficient for the calibration of wide-field instruments, as demonstrated in Paper~I for WALOPs and by \cite{moonlight_calibration_VLT} for the FOcal Reducer and low dispersion Spectrograph (FORS2) polarimeter (described later).

One promising candidate for polarimetric flat fields is the sky on bright-Moon nights \citep{moonlight_calibration_VLT}. During such times, owing to the geometry of the Sun-Earth-Moon system, the light entering the atmosphere from the Moon is unpolarized on full-Moon nights or polarized up to a low level when within a few days of it.

While traversing the atmosphere, the polarization state of the light beam is modified primarily due to the scattering by small atmospheric molecules, described by Rayleigh scattering. Therefore, the observed polarization depends on the scattering geometry between the observer (telescope), the sky location, and the position of the Moon in the sky.
Assuming that the atmosphere can be described by a single-layer scattering region, Rayleigh scattering predicts that, for unpolarized light on full-Moon nights, the polarization fraction, $p$, depends on the angular distance of the region, $\gamma$, from the Moon, as given by Eq.~\ref{moonlight_p_eqn1} (\citealt{moon_pol_ref}; \citealt{moon_pol_ref2}; \citealt{Harrington_solar_Calibration}; \citealt{Strutt_moonlight_polarization}; \citealt{sky_polarization_reference_0}).
\begin{equation}\label{moonlight_p_eqn1}
p = \delta\frac{\sin^{2}\gamma}{1 + \cos^{2}\gamma},
\end{equation}
$\delta$ is an empirical parameter whose value depends on the sky conditions, and for clear cloudless nights, it is found to be around 0.8 \citep{moon_pol_ref}.

\begin{figure}
    \centering
    \includegraphics[trim={0cm 0cm 1.2cm 0cm},clip,width=.98\columnwidth]{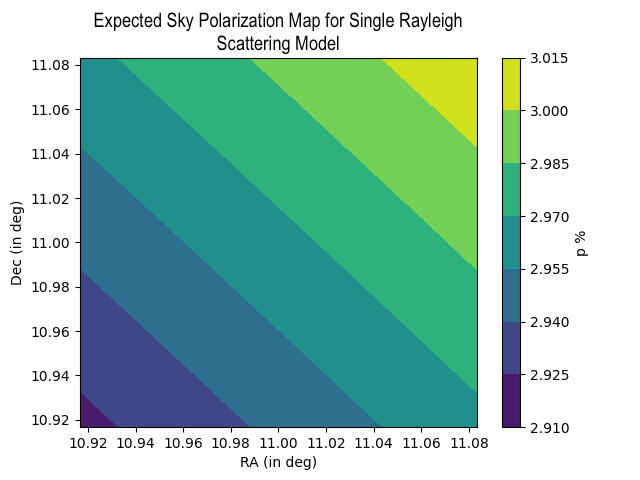}\\[-1.5ex]
    \caption{Simulated polarization for a $10\times10$~arcminutes patch at a distance of $15.5^\circ$ from the Moon based on a single-layer Rayleigh scattering model to describe the atmosphere. In this example, the coordinates of the Moon are (RA = $0^{\circ}$, Dec = $0^{\circ}$), while the patch is centered at (RA = $11.0^{\circ}$, Dec = $11.0^{\circ}$) in the GCRS coordinate system. }
    \label{sky_polarization_expected}
\end{figure}

As can be seen, the value of $p$ increases as the pointing ($\gamma$) is farther away from the Moon, with the maximum value at $\gamma = \pi/2$, whereas it is zero near the vicinity of the Moon. The expected Electric Vector Position Angle (EVPA), $\theta$, is a function of the sky position of the Moon as well as the sky pointing. By choosing a suitable combination of these, desired EVPAs can be 
 achieved. This way, required combinations of $p$ and $\theta$ (i.e., $q$ and $u$) can be obtained depending on the calibration requirements.

This scattering model 
predicts that within an area of $10\times10$~arcminutes and sky positions of up to 15-20 degrees away from the Moon, $p$ (and, $q$ and $u$), will remain constant to a level of few hundredths of a percent (see Fig.~\ref{sky_polarization_expected}).
 While deviations from this simple picture may arise due to several complicating factors, the polarization is still likely to remain constant within such a patch.
Previously, assuming the sky on the full-Moon nights as a linear polarimetric flat source, \cite{moonlight_calibration_VLT} calibrated the FORS2 polarimeter mounted at the Very Large Telescope (VLT), which has a FoV of 7~arcminutes, to an accuracy better than 0.05~\% in $p$. They assumed the instrumental polarization to be zero at the center of the FoV and used the measured linear polarization there as the polarization of the sky across the FoV.

In this work, we carried out linear polarimetric observations of the extended sky greater than ten arcminutes in size to verify the suitability of the polarized sky as a wide-field polarimetric flat source. We used the RoboPol instrument to observe a total of 19 patches on five different nights within two days of the full-Moon in the SDSS-r band filter. Section~\ref{observations} presents the details of the observations carried out for this study. The data analysis is presented in Sect.~\ref{analysis} where we find that 12 of the 19 patches are simultaneously uniform in $q$ and $u$ to within 0.07~\%, while the other patches are uniform to within 0.3~\%. We discuss our results in Sect.~\ref{discussion} and provide our conclusions and an outlook for future works in Sect.~\ref{conclusions}.

\section{Observations}\label{observations}

\begin{figure}
    \centering
    \includegraphics[trim={0.2cm 1.cm 0cm 1.5cm},clip,width=.98\columnwidth]{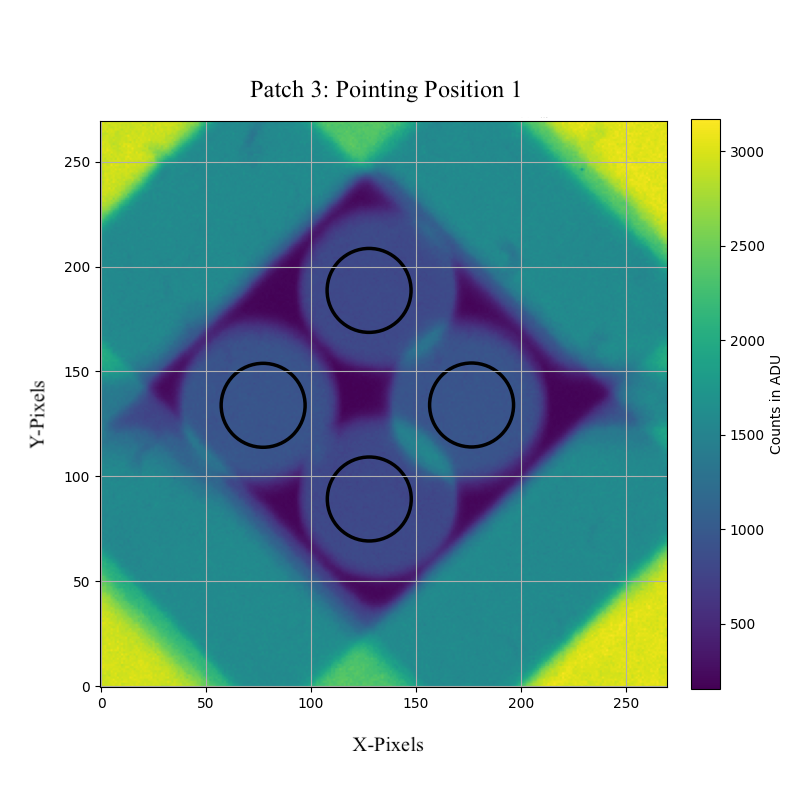}\\[-1.5ex]
    \caption{Image of the central region of RoboPol used for making the measurements. Four images of the source, corresponding to $0^{\circ}$, $45^{\circ}$, $90^{\circ}$ and $135^{\circ}$ polarization are formed simultaneously at the CCD and differential photometry of the two pairs of images yields the Stokes parameters $q$ and $u$. The apertures used for the photometry of the four channels are marked in black circles. The color bar indicates the counts in ADUs for the exposure.}
    \label{rbpl_mask}
\end{figure}

\begin{figure}
    \begin{subfigure}{0.49\textwidth}
    \centering
    \includegraphics[scale = 0.52]{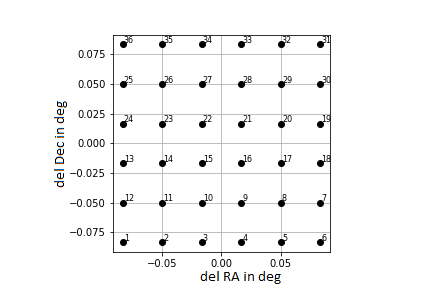}
    \caption{Nominal pointing positions and their sequence for Patches 1 to 11. The overall size of the patch is $10 \times 10$~arcminutes.}
    \label{nominal_pos}
    \end{subfigure}
    \begin{subfigure}{0.49\textwidth}
    \centering
    \includegraphics[scale = 0.5]{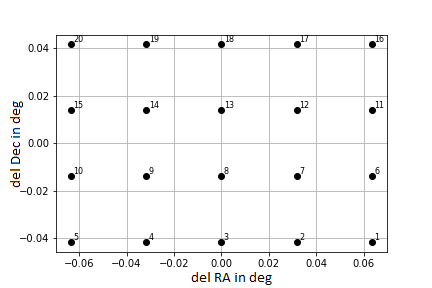}
    \caption{Nominal pointing positions  and their sequence for Patches 12 to 19. The overall size of the patch is $4.8\times 7.5$~arcminutes.}
    \label{nominal_pos2}
    \end{subfigure}
    \begin{subfigure}{0.49\textwidth}
    \centering
    \includegraphics[scale = 0.42]{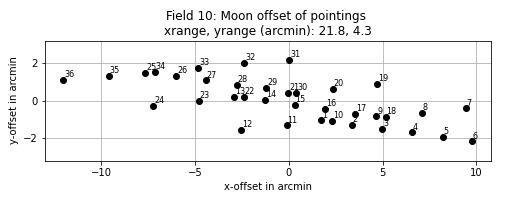}
    \caption{Observed pattern of the position of the grid points with respect to the Moon for Patches 1-11. For Patch 10 shown here, the grid points are spread across a region of $21.8 \times 4.3$~arcminutes with respect to the geometrical center.}
    \label{actual_pos1}
    \end{subfigure}
    \begin{subfigure}{0.49\textwidth}
    \centering
    \includegraphics[scale = 0.42]{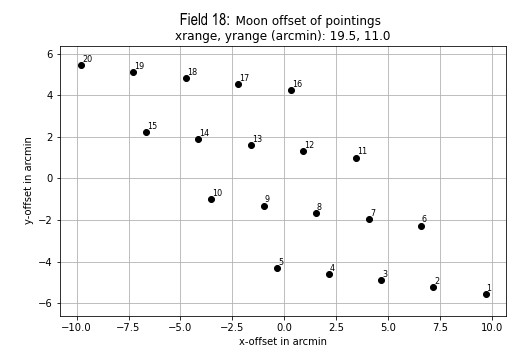}
    \caption{Observed pattern of the position of the grid points with respect to the Moon for Patches 12-19. For Patch 18 shown here, the grid points are spread across a region of $19.5 \times 11$~arcminutes with respect to the geometrical center.}
    \label{actual_pos2}
    \end{subfigure}

\caption{Sky patches of size $10\times10$ (Patches 1-11) arcminutes and $4.8\times7.5$ arcminutes (Patches 12-19) were observed with RoboPol through a rectangular grid of either $6 \times 6$ or $5 \times 4$ pointing coordinates, respectively, in the sequence marked in the images. Due to the motion of the Moon in the sky, the shape of the grid with respect to the Moon is distorted, as shown in (c) and (d), depending on the sequence of observations followed.}
\label{pointing_grids}
\end{figure}

\begin{table*}
    \begin{center}
    \caption{\label{tab:obs1} Details of observed sky patches.}
\begin{tabular}{ccccccccc}
\hline \hline \\ [-1.5ex]
  \begin{tabular}[c]{@{}c@{}}Patch\\ \# \\ \\ \end{tabular} &
  \begin{tabular}[c]{@{}c@{}}RA\\ {[}deg{]}\\ \\ \end{tabular} &
  \begin{tabular}[c]{@{}c@{}}Dec\\ {[}deg{]}\\ \\ \end{tabular} &
  \begin{tabular}[c]{@{}c@{}}Moon\\ RA\\ {[}deg{]} \\ \end{tabular} &
  \begin{tabular}[c]{@{}c@{}}Moon\\ Dec\\ {[}deg{]} \\ \end{tabular} &
  \begin{tabular}[c]{@{}c@{}}Moon
  \\ Dist.\\ {[}deg{]} \\  \end{tabular} &
  \begin{tabular}[c]{@{}c@{}}Patch\\ size\\ {[}arc-min{]} \\ \end{tabular} &
  \begin{tabular}[c]{@{}c@{}}Days\\ after\\ full-Moon \\ \end{tabular} &
  \begin{tabular}[c]{@{}c@{}}Observation\\ Date \\
  {[}dd-mm-yyyy{]} \\  \end{tabular} \\[+.5ex] \hline\\[-1.5ex]
1                      & 350.0                                                                                   & 3.8                                                                                      & 10.6                                                                                           & -0.5                                                                                                         & 21                                                      & $40 \times 24$                        & \multirow{3}{*}{1}                                                              & \multirow{3}{*}{21-09-2021}                                                                       \\
2                      & 0.8                                                                                     & 4.1                                                                                      & 11.2                                                                                           & 0.0                                                                                                          & 11.2                                                    & $24 \times 5$                         &                                                                                 &                                                                                                   \\
3                      & 15.0                                                                                    & 0.0                                                                                      & 11.5                                                                                           & 0.2                                                                                                          & 3.5                                                     & $21 \times 3$                         &                                                                                 &                                                                                                   \\[+.5ex] \hline\\[-1.5ex]
4                      & 25.4                                                                                    & 15.2                                                                                     & 21.9                                                                                           & 4.9                                                                                                          & 10.9                                                    & $21 \times 4$                         & \multirow{4}{*}{2}                                                              & \multirow{4}{*}{22-09-2021}                                                                       \\
5                      & 20.0                                                                                    & -2.9                                                                                     & 22.3                                                                                           & 5.2                                                                                                          & 8.4                                                     & $23 \times 2.5$                         &                                                                                 &                                                                                                   \\
6                      & 8.0                                                                                     & 5.5                                                                                      & 22.6                                                                                           & 5.5                                                                                                          & 14.5                                                    & $25 \times 5$                        &                                                                                 &                                                                                                   \\
7                      & 30.1                                                                                    & -2.5                                                                                     & 22.9                                                                                           & 5.7                                                                                                          & 10.9                                                    & $23 \times 2$                         &                                                                                 &                                                                                                   \\[+.5ex] \hline\\[-1.5ex]
8                      & 39.8                                                                                    & 11.2                                                                                     & 41.7                                                                                           & 14.1                                                                                                         & 3.4                                                     & $20 \times 5$                         & \multirow{4}{*}{1}                                                              & \multirow{4}{*}{21-10-2021}                                                                       \\
9                      & 36.7                                                                                    & 15.8                                                                                     & 41.9                                                                                           & 14.3                                                                                                         & 5.3                                                     & $22 \times 4.5$                       &                                                                                 &                                                                                                   \\
10                     & 40.0                                                                                    & 9.2                                                                                      & 42.2                                                                                           & 14.4                                                                                                         & 5.6                                                     & $22 \times 4.3 $                      &                                                                                 &                                                                                                   \\
11                     & 40.0                                                                                    & 7.0                                                                                      & 42.5                                                                                           & 14.5                                                                                                         & 7.9                                                     & $21 \times 4.3 $                      &                                                                                 &                                                                                                   \\[+.5ex] \hline\\[-1.5ex]
12                     & 213.8                                                                                   & 0.0                                                                                      & 213.9                                                                                          & -12.7                                                                                                        & 12.7                                                    & $19 \times 13.7$                      & \multirow{2}{*}{-1}                                                             & \multirow{2}{*}{14-05-2022}                                                                       \\
13                     & 220.8                                                                                   & -5                                                                                       & 214.3                                                                                         & -12.9                                                                                                       & 10.2                                                    & $19 \times 13.3$                     &                                                                                 &                                                                                                   \\[+.5ex] \hline\\[-1.5ex]
14                     & 228.3                                                                                   & -15                                                                                      & 228.7                                                                                         & -18.7                                                                                                       & 3.8                                                     & $17 \times 10 $                       & \multirow{6}{*}{0}                                                              & \multirow{6}{*}{15-05-2022}                                                                       \\
15                     & 228.6                                                                                   & -10.3                                                                                    & 228.3                                                                                         & -18.5                                                                                                       & 8.2                                                     & $16 \times 11.5 $                     &                                                                                 &                                                                                                   \\
16                     & 236                                                                                     & -14.7                                                                                    & 228.9                                                                                         & -18.8                                                                                                       & 8                                                       & $18.5 \times 10.9$                    &                                                                                 &                                                                                                   \\
17                     & 220.8                                                                                   & -9.9                                                                                     & 228.0                                                                                         & -18.3                                                                                                       & 11                                                      & $20 \times 12.8 $                     &                                                                                 &                                                                                                   \\
18                     & 235.5                                                                                   & -10                                                                                      & 229.1                                                                                         & -19.0                                                                                                       & 10.9                                                    & $19.5 \times 11.0 $                   &                                                                                 &                                                                                                   \\
19                     & 221.6                                                                                   & -14.9                                                                                    & 228.5                                                                                         & -18.6                                                                                                       & 7.6                                                     & $18.5 \times 11.5 $                   &                                                                                 &                                                                                                  \\[+.5ex] \hline
\end{tabular}
 \tablefoot{The coordinates of the Moon and the sky patches are given in Geocentric Coordinate Reference System (GRCS) based equatorial coordinate system. All observations were spread over 5 nights and two observation seasons. RA and Dec are the central (mean) coordinates, and Patch Size is the overall extent of the patch accounting for the Moon's motion in the sky during the observations.}
 \end{center}
\end{table*}

\par All observations were performed in the SDSS-r filter with the RoboPol instrument mounted on the 130\,cm~Telescope of the Skinakas Observatory in Crete, Greece. The instrument is described in detail by \citet{robopol}. It is a four-channel one-shot optical linear polarimeter that measures $q$ and $u$ in a single exposure. The four channels are projected on the same CCD, as shown in Fig.~\ref{rbpl_mask}. A central mask blocks light from neighboring regions of the observed target field, increasing the accuracy by reducing the sky background. We observed 19~patches of the sky at different separations from the Moon and during different Moon phases, as listed in Table~\ref{tab:obs1}. Each patch nominally covered an area of either $10 \times 10$ (Patches 1-11) or $5 \times 7.5$~arcminutes (Patches 12-19) in the geocentric celestial reference system (GCRS) based equatorial coordinates, as shown in Figs.~\ref{nominal_pos} and~\ref{nominal_pos2}. The coordinates of the Moon and the sky presented in this work and used for calculations are in the GCRS as the motion of the Moon is bound by Earth's gravity. Each patch was divided into a rectangular grid of points (coordinates) which were observed through RoboPol inside the mask. The observed patches were divided into a grid of $6 \times 6$ pointings during the first 3 nights (Patches 1-11) and into a grid of $5 \times 4$ pointings for the remaining observations (Patches 12-19).

\par The observation sequence of the grid points for a patch was chosen keeping the following two effects into consideration: 
\begin{itemize}
    \item While the observations of any patch are being carried out, the sky position of the Moon changes with respect to the patch due to its non-sidereal motion. So the rectangular grid becomes distorted with respect to the Moon.
    \item The overall accuracy of the telescope pointing is 2 arcminutes if the telescope slews (when the separation between consecutive grid points is greater than 8 arcminutes). Whereas, if the separation is less than 8 arcminutes, the telescope moves through very precise and small offset motion with an accuracy of a few arcseconds, yielding high accuracy pointing.

\end{itemize}

Thus, the observation sequence for the grid points of a patch was decided so as to minimize the effect of the Moon's motion on the patch size and morphology, yet at the same time using only small offsets to move the telescope and obtain high accuracy pointing. Two different observation sequences were followed for different nights (Figs.~\ref{nominal_pos} and~\ref{nominal_pos2}), leading to different morphologies of the grid points of the patches with respect to the Moon. For Patches 1-11, the square grid becomes distorted with respect to the Moon resembling a pattern as shown in Fig.~\ref{actual_pos1}. Figure~\ref{actual_pos2} shows the corresponding plot for Patches 12-19. The patch size noted in Table~\ref{tab:obs1} is the overall extent in the Right ascension (RA) and declination (Dec) coordinates for the patches. As can be seen, the grid points which were nominally spread over up to $10 \times 10$~arcminutes in form of a rectangular grid become distorted and spread over tens of arcminutes on the sky with respect to the Moon. 
\par  The exposure time per pointing ranged from 2.5~to~90~seconds and was chosen for each patch such that the uncertainty in measured fractional polarization from photon noise was 0.04-0.05\% in the central masked region. The exposure time depends on the sky's brightness, which itself is a function of the angular separation from the Moon as well as the Moon phase. Furthermore, patches for the observation of the sky were chosen using the following two criteria: (a) to sample various distances as well as orientations with respect to the Moon, and (b) the patch should contain very few stars (and no bright stars). During the observations, it was ensured that no star fell inside the central masked region.
Multiple standard stars, used to calibrate the RoboPol polarimeter, were observed during the observation nights. The instrumental zero polarization obtained from those measurements was consistent with the values reported in Table~\ref{tab:inst_pol}, obtained for the full observing seasons following the standard procedure (\citealt{robopol_data_release}, Blinov et al. 2023, \textit{in prep.}).

\section{Results}\label{analysis}

A dedicated data reduction pipeline was written in Python to analyze the raw data. Aperture photometry (without any background subtraction) was carried out using the Photutils package (\citealt{photutils}) on the images to obtain the intensities of the four beams of the sky on the CCD (Fig.~\ref{rbpl_mask}).
Circular apertures of size 12 arcseconds were used.
From these, the Stokes parameters were found using the normalized difference between corresponding intensities.
The instrumental zero polarization was then subtracted from these measurements.
Throughout the analysis, careful attention was given to the error estimation and propagation in each step.

\begin{table}
    \centering
    \caption{\label{tab:inst_pol} Instrumental zero polarization of RoboPol during the observation runs (Blinov et al. 2023, \textit{in prep.}).}
    \begin{tabular}{ccc}
        \hline \hline \\ [-1.5ex]
        Observation Run & $q_{\rm{inst}}$ & $u_{\rm{inst}}$ \\
        & [\%] & [\%] \\[+.5ex]
        \hline \\[-1.ex]
         Sep, Oct 2021 & 0.25 $\pm$ 0.15 & - 0.36 $\pm$ 0.09\\
         May 2022 & 0.50 $\pm$ 0.12 & - 0.37 $\pm$ 0.08\\[+.5ex]
         \hline
    \end{tabular}
    \label{inst_pol}
\end{table}

\par To check for the polarimetric flatness of a patch, we calculated the mean and the standard deviation of the normalized Stokes parameters $q$ and $u$ for all the $n$ grid points using the conventional formulae, as shown in Eqs.~\ref{q_mean}, \ref{u_mean}, \ref{q_std} and~\ref{u_std}.

\begin{equation}\label{q_mean}
    q = q_{mean} = \frac{1}{n} \sum_{i=1}^{n} q_{i}=\frac{1}{n}\left(q_{1}+\cdots+q_{n}\right)
\end{equation}

\begin{equation}\label{u_mean}
    u = u_{mean} = \frac{1}{n} \sum_{i=1}^{n} u_{i}=\frac{1}{n}\left(u_{1}+\cdots+u_{n}\right)
\end{equation}

\begin{equation}\label{q_std}
    \sigma_{q}^{2} = \frac{1}{n-1} \sum_{i=1}^{n} (q_{i} - q_{mean})^{2}
\end{equation}

\begin{equation}\label{u_std}
    \sigma_{u}^{2}  = \frac{1}{n-1} \sum_{i=1}^{n} (u_{i} - u_{mean})^{2}
\end{equation}

\begin{figure}
     \centering
    \includegraphics[scale = 0.4]{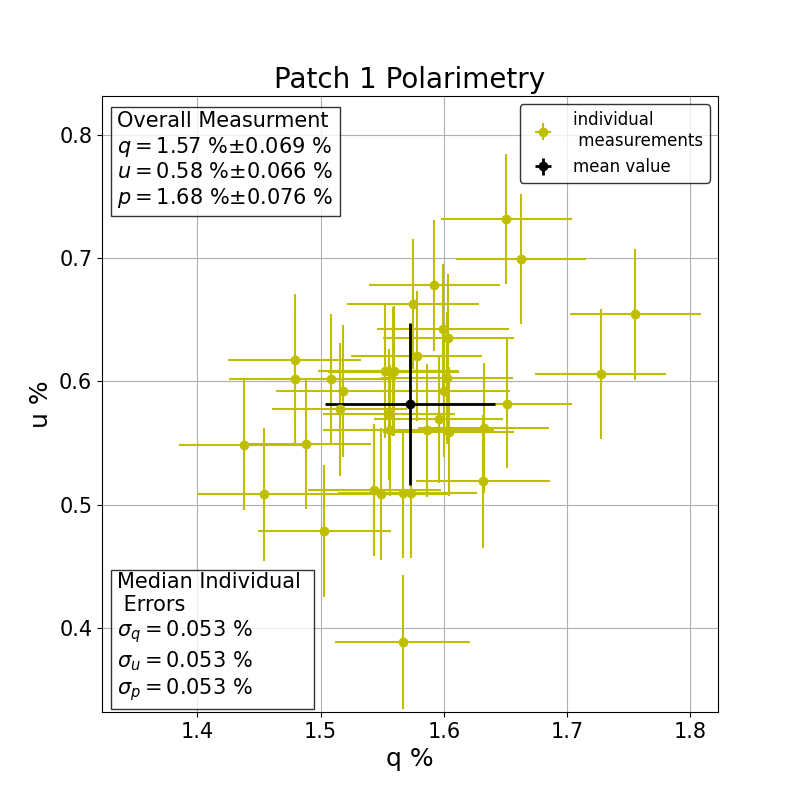}
\caption{Plot of measured $q$ and $u$ values for all the pointings for Patch 1 (yellow crosses) as well as the overall mean value and the measured standard deviation (black cross), whose values are mentioned in the top left legend. The median individual errors in the bottom left legend refer to the median of expected errors in the Stokes parameters owing to photon noise across the grid points. The value of the degree of linear polarization, $p$, is presented without any correction for polarimetric bias. $p$ is calculated for each grid point and then their mean and variance are found, similar to $q$ and $u$.}
\label{qu_maps}
\end{figure}

Figure~\ref{qu_maps} shows the measured polarizations in the $q-u$ plane for Patch~1. Corresponding plots for all the other patches are shown in Figs.~\ref{fig_app_1} and~\ref{fig_app_2}. These measurements of polarimetric flatness for all the patches are listed in Table~\ref{tab:results_short}. In general, we find the patches to have a scatter in $q$ and $u$ under 0.07~\%. For 12 of the 19 patches, we found $q$ and $u$ both to be constant within $0.07~\%$, with the maximum value reaching 0.30~\% for Patch~16. We find a higher spread ($>= 0.1~\%$) in the measurement of either $q$ and/or $u$ in Patches~13, 15, 16, 17, and~19.

We draw attention to the fact that the correction of the instrumental polarization on the measurements does not affect the estimates of the standard deviation in the patches, and thus of the flatness of the sky polarization. The observational quantification of these dispersions is the main result of this paper, as it confirms that the sky polarization around the full-Moon is a good flat-source candidate in sky regions of 10-by-10 arcminutes or more.

\section{Discussion}\label{discussion}
\begin{table}
\begin{center}
\caption{\label{tab:results_short} Mean and standard deviation of linear polarization properties in patches.}
\begin{tabular}{cccccccc}
\hline \hline \\ [-1.5ex]
Patch & $\gamma$ & $q$ & $\sigma_q$ & $u$ & $\sigma_u$ & $p$ & $\sigma_p$ \\
\# & [deg] & [\%] & [\%] & [\%] & [\%] & [\%] & [\%]
   \\[+.5ex]\hline\\[-1.5ex]
1  & 21   & 1.57  & 0.07 & 0.58  & 0.07 & 1.68 & 0.08 \\
2  & 11.2 & 0.02  & 0.09 & 0.1   & 0.06 & 0.14 & 0.06 \\
3  & 3.5  & -0.24 & 0.04 & -0.06 & 0.05 & 0.25 & 0.04 \\
4  & 10.9 & 0.14  & 0.05 & -1.01 & 0.07 & 1.02 & 0.07 \\
5  & 8.4  & 0.2   & 0.05 & -0.36 & 0.04 & 0.41 & 0.05 \\
6  & 14.5 & 1.25  & 0.07 & 0.26  & 0.05 & 1.28 & 0.07 \\
7  & 10.9 & 0.43  & 0.05 & 0.74  & 0.06 & 0.85 & 0.06 \\
8  & 3.4  & 0.14  & 0.05 & 0.24  & 0.06 & 0.28 & 0.05 \\
9  & 5.3  & 0.43  & 0.05 & 0.12  & 0.06 & 0.44 & 0.06 \\
10 & 5.6  & -0.08 & 0.07 & -0.23 & 0.05 & 0.25 & 0.05 \\
11 & 7.9  & -0.65 & 0.07 & 0.05  & 0.07 & 0.66 & 0.07 \\
12 & 12.7 & -1.0  & 0.07 & -0.46 & 0.05 & 1.11 & 0.05 \\
13 & 10.2 & -0.6  & 0.16 & -1.7  & 0.04 & 1.81 & 0.04 \\
14 & 3.8  & -0.58 & 0.05 & -0.04 & 0.05 & 0.59 & 0.05 \\
15 & 8.2  & -2.03 & 0.07 & 0.21  & 0.1  & 2.05 & 0.08 \\
16 & 8.0  & 0.52  & 0.06 & -1.32 & 0.3  & 1.42 & 0.28 \\
17 & 11.0 & -1.06 & 0.06 & 0.8   & 0.21 & 1.34 & 0.12 \\
18 & 10.9 & -0.75 & 0.07 & -1.95 & 0.08 & 2.09 & 0.07 \\
19 & 7.6  & 0.39  & 0.18 & 0.88  & 0.07 & 0.97 & 0.12 \\[+.5ex]
\hline
\end{tabular}
\tablefoot{The zero instrumental polarization (Table~\ref{tab:inst_pol}) has been removed from the measurements. The degree of polarization have not been corrected for polarization biases.}
\end{center}
\end{table}

As shown in Table~\ref{tab:results_short}, the standard deviation within each patch is typically less than 0.07~\%. Several effects contribute to the observed scatter of measurements in individual patches:
photon noise, variability in the instrumental polarization, the gradient in sky polarization as a function of distance, and possibly, changes in sky polarization during the observations. 

As already noted, the exposure times during observations were adjusted for each patch such that the achieved photon noise contribution is around 0.05~\% for all our $q$ and $u$ measurements.

The stability of RoboPol is around 0.15~\% over an entire observation season from 2014-2022 \citep{robopol_data_release}.
However, for the observations presented in this work, we find that the instrumental polarization is non-variable to within 0.07~\% during the observations of each patch,
on the timescale of half an hour to two hours. These low values indicate that the change in instrumental polarization is small for such exposure times and for small sky regions, thus mitigating any source of systematic from possible instrumental flexure and other sources of variable instrumental polarization. 

Another contribution to the scatter is the fact that the polarization is expected to change depending on the position within the patch, according to Eq.~\ref{moonlight_p_eqn1}, and as shown in Fig.~\ref{sky_polarization_expected}. Our results demonstrate that this effect is less than 0.1\% levels for patches extending up to 20~arcminutes.

Finally, we must notice that the sky polarization may change during the observations due to fluctuations in atmospheric conditions. While it might be the dominant source of scatter in our measurements for Patches~13, 15, 16, 17, and~19, our observations show that this source of scatter remains lower than 0.3\% (in $p$) within the time scales needed to observe individual patches (half an hour to two hours).
A wide-field polarimeter like WALOP will observe an entire patch in a single exposure of a much shorter time than required with RoboPol. Therefore, we expect that the time variability of the sky polarization will not play a significant role.

\section{Conclusions}\label{conclusions}
Currently, no known and established polarimetric flat sources exist for the calibration of wide-field optical polarimeters like WALOPs. A critical ingredient in the on-sky calibration method of the WALOP polarimeters is the use of multiple partially polarized polarimetric flat sources whose polarization values are spread across the $q-u$ plane.

In this paper, we have experimentally demonstrated that the sky in the vicinity of the full-Moon can be used as an extended linear polarimetric flat source for the relative calibration of wide-field linear polarimeters. The sky polarization indeed remains constant at the level of 0.1\% or lower in sky regions of 10~to 20~arcminutes. Furthermore, different combinations of $q-u$ can be achieved based on the relative sky positions of the Moon and the target patch.

While we have only demonstrated this in SDSS-r band and for the bright-Moon sky within 2~days of the full-Moon, it is expected to hold true for other filters in the optical wavelengths as the polarizing mechanism remains the same. In the near future, we plan to carry out similar measurements with RoboPol in other broadband filters to confirm this.

\begin{acknowledgements}
The \textsc{Pasiphae} program is supported by grants from the European Research Council (ERC) under grant agreement No 771282 and No 772253, from the National Science Foundation, under grant number AST-1611547 and the National Research Foundation of South Africa under the National Equipment Programme. This project is also funded by an infrastructure development grant from the Stavros Niarchos Foundation and from the Infosys Foundation. VPa acknowledges support by the Hellenic Foundation for Research and Innovation (H.F.R.I.) under the “First Call for H.F.R.I. Research Projects to support Faculty members and Researchers and the procurement of high-cost research equipment grant” (Project 1552 CIRCE), and from the Foundation of Research and Technology - Hellas Synergy Grants Program through project MagMASim, jointly implemented by the Institute of Astrophysics and the Institute of Applied and Computational Mathematics. KT acknowledges support from the Foundation of Research and Technology - Hellas Synergy Grants Program through project POLAR, jointly implemented by the Institute of Astrophysics and the Institute of Computer Science. This work was supported by NSF grant AST-2109127. SM would like to thank Anna Steiakaki for providing careful comments and corrections to the various drafts of the manuscript.
\\
\par This work utilized the open source software packages Astropy (\cite{astropy:2013, astropy:2018}), Numpy (\cite{numpy}, Scipy (\cite{scipy}), Matplotlib (\cite{matplotlib}) and Jupyter notebook (\cite{jupyter_notebook2}).
\end{acknowledgements}

\bibliographystyle{aa}
\bibliography{references} 

\appendix

\onecolumn
\section{Polarimetric Measurement Plots of all Patches}

\begin{figure}
\begin{subfigure}{0.33\textwidth}
     \centering
    \includegraphics[scale = 0.3]{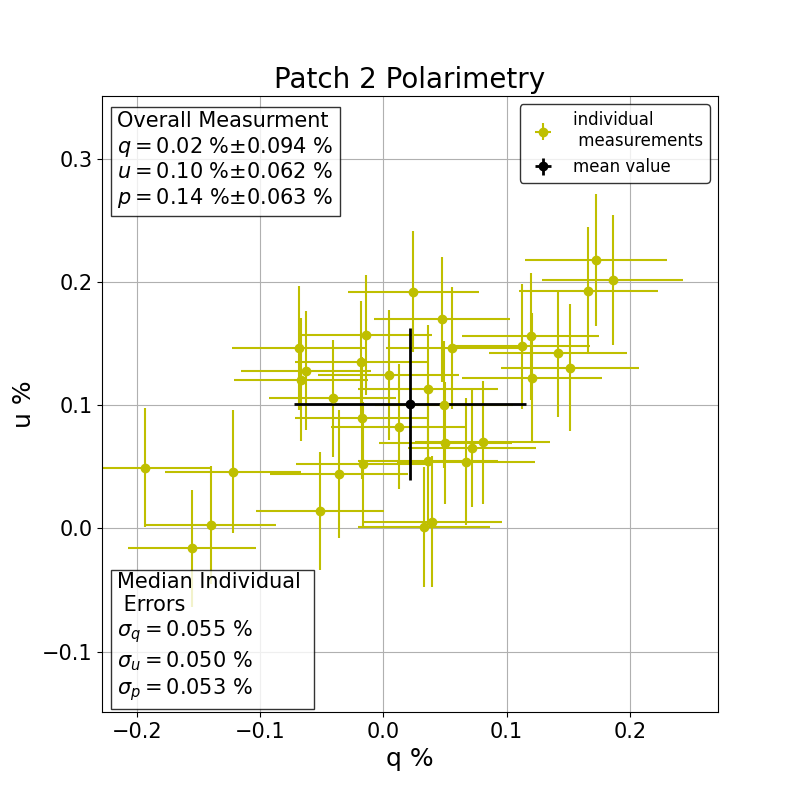}
    \caption{Patch 2}
    \label{qu2}
\end{subfigure}
\begin{subfigure}{0.33\textwidth}
     \centering
    \includegraphics[scale = 0.3]{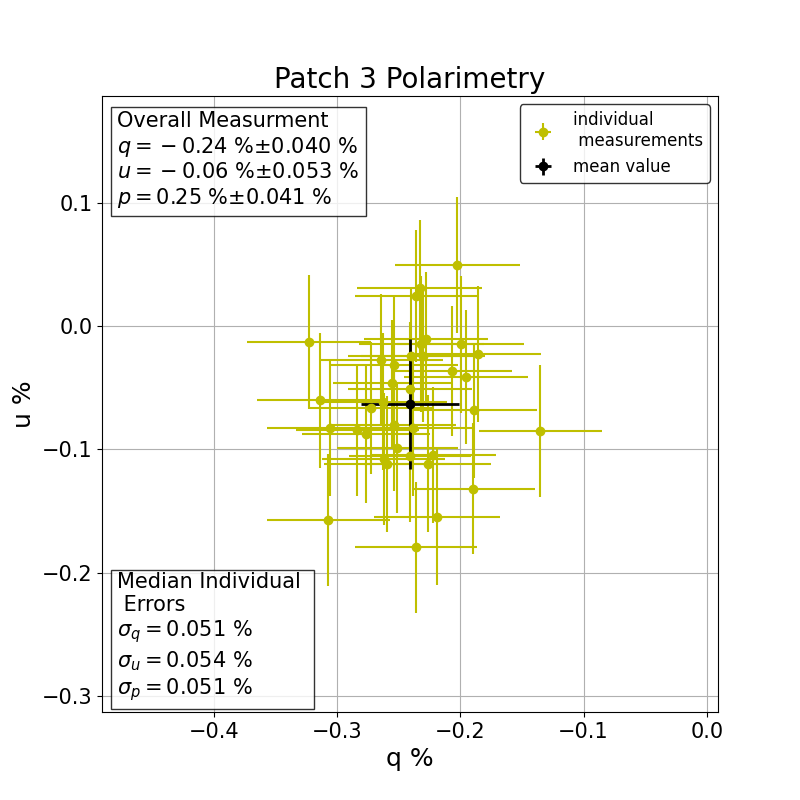}
    \caption{Patch 3}
    \label{qu3}
\end{subfigure}
\begin{subfigure}{0.33\textwidth}
     \centering
    \includegraphics[scale = 0.3]{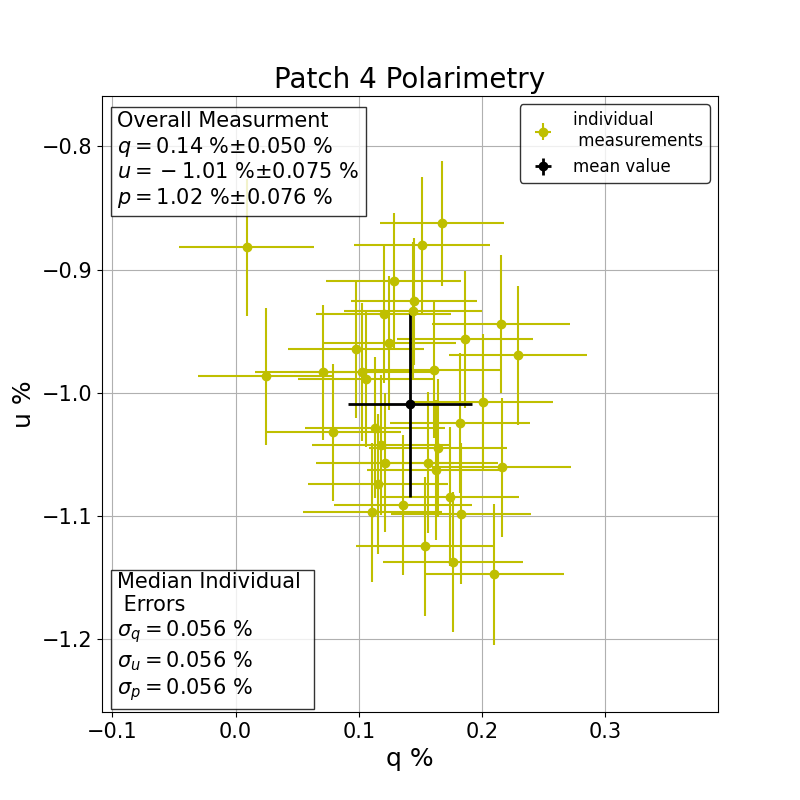}
    \caption{Patch 4}
    \label{qu4}
\end{subfigure}
\begin{subfigure}{0.33\textwidth}
     \centering
    \includegraphics[scale = 0.3]{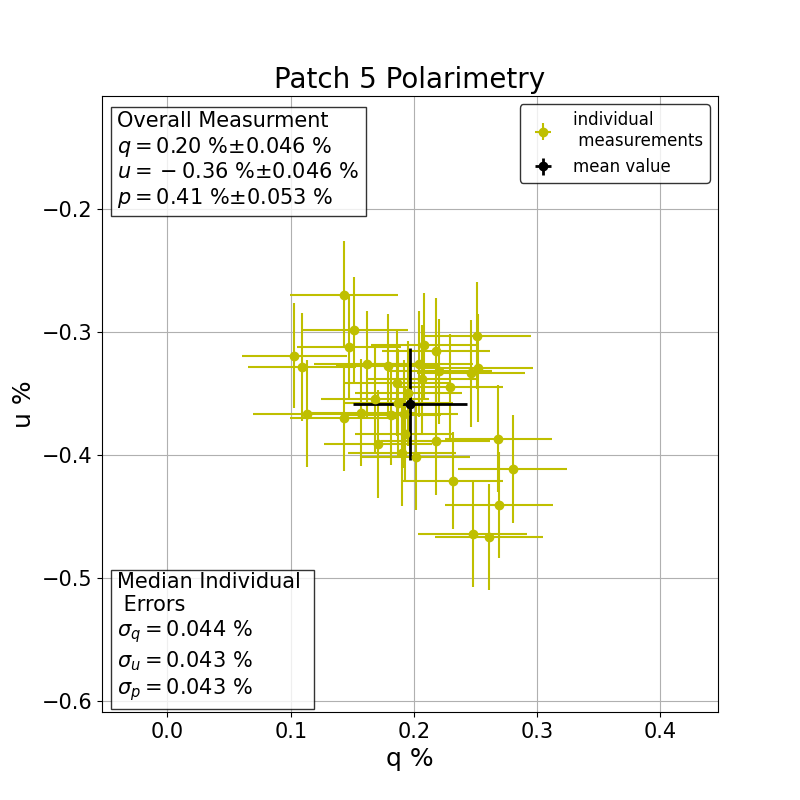}
    \caption{Patch 5}
    \label{qu5}
\end{subfigure}
\begin{subfigure}{0.33\textwidth}
     \centering
    \includegraphics[scale = 0.3]{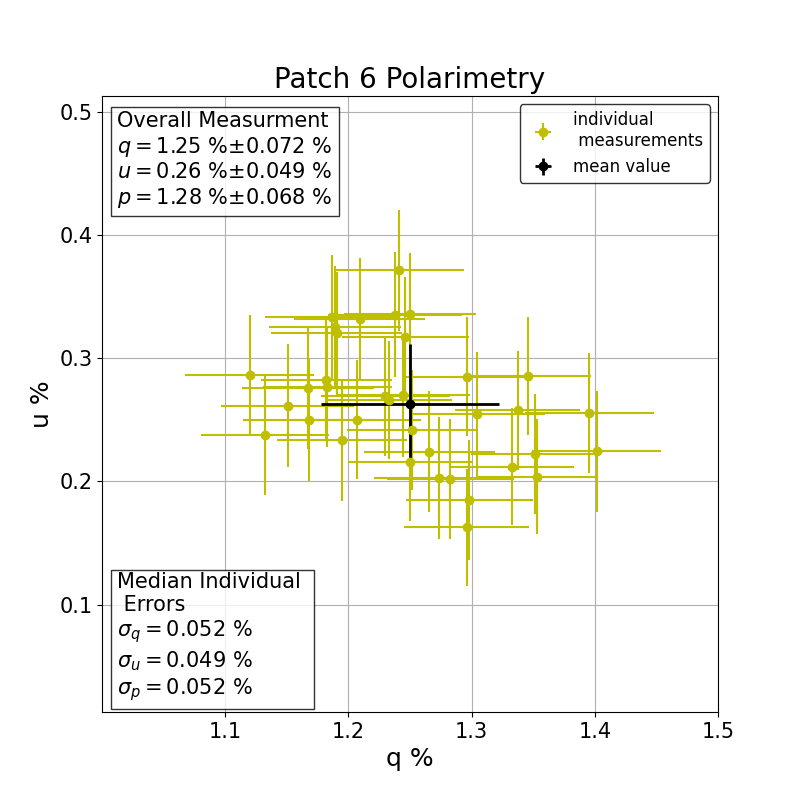}
    \caption{Patch 6}
    \label{qu6}
\end{subfigure}
\begin{subfigure}{0.33\textwidth}
     \centering
    \includegraphics[scale = 0.3]{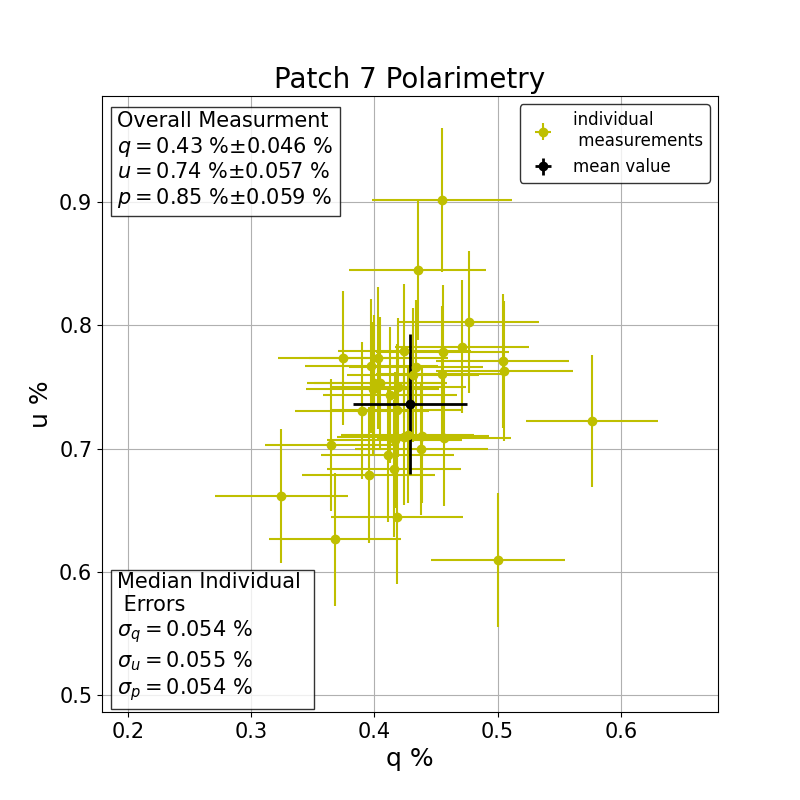}
    \caption{Patch 7}
    \label{qu7}
\end{subfigure}
\begin{subfigure}{0.33\textwidth}
     \centering
    \includegraphics[scale = 0.3]{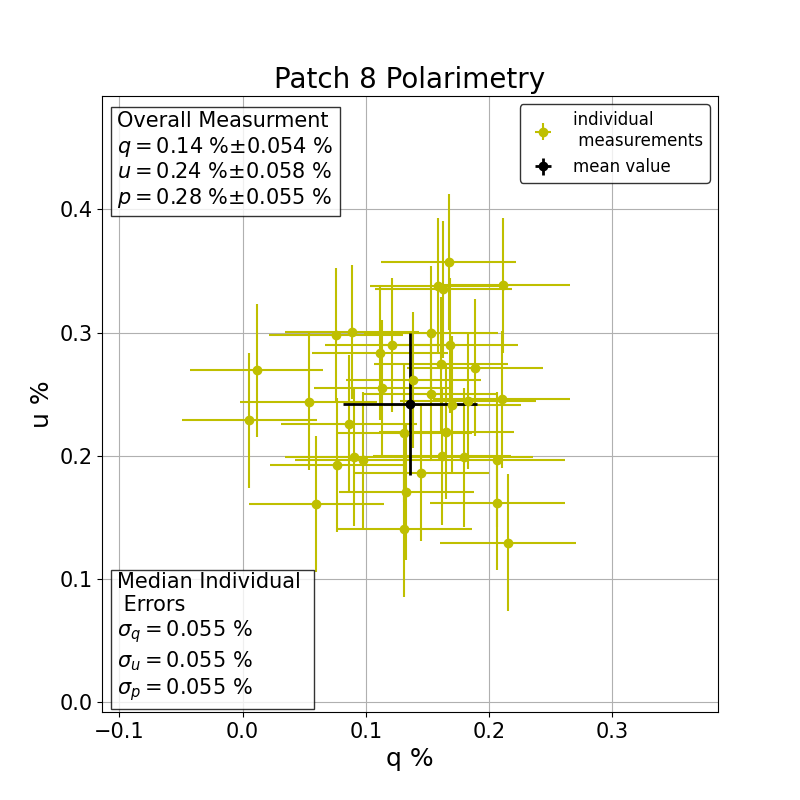}
    \caption{Patch 8}
    \label{qu8}
\end{subfigure}
\begin{subfigure}{0.33\textwidth}
     \centering
    \includegraphics[scale = 0.3]{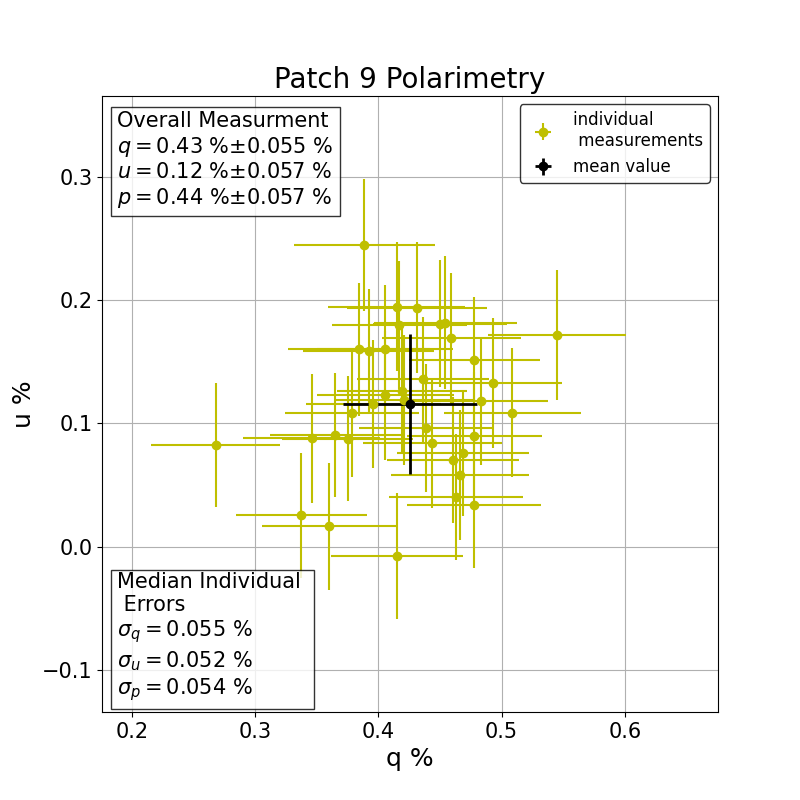}
    \caption{Patch 9}
    \label{qu9}
\end{subfigure}
\begin{subfigure}{0.33\textwidth}
     \centering
    \includegraphics[scale = 0.3]{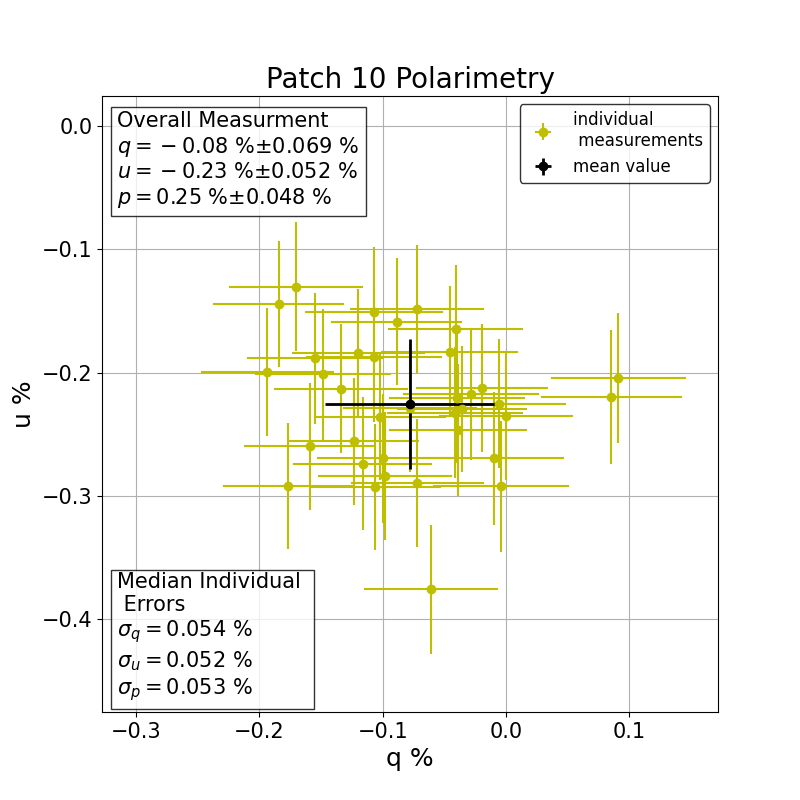}
    \caption{Patch 10}
    \label{qu10}
\end{subfigure}

\caption{Polarimetric measurements for Patches 2 to 10, similar to Fig.~\ref{qu_maps}. Yellow crosses are $q$ and $u$ values for all the individual pointings for a  patch. The overall mean value and the measured standard deviation (black cross) are mentioned in the top left legend.}
\label{fig_app_1}
\end{figure}

\begin{figure}
\begin{subfigure}{0.33\textwidth}
     \centering
    \includegraphics[scale = 0.3]{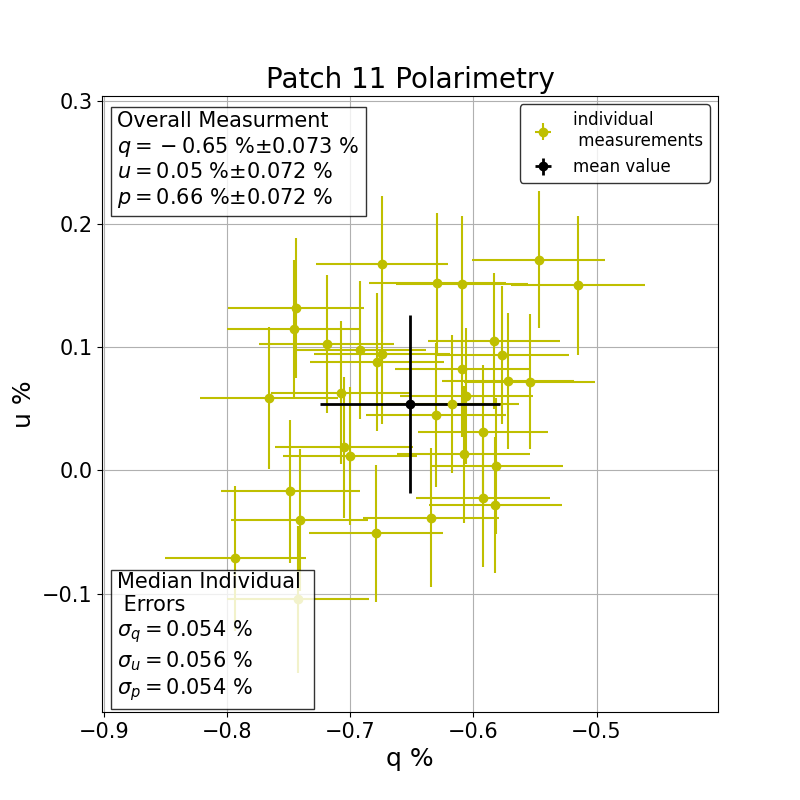}
    \caption{Patch 11}
    \label{qu11}
\end{subfigure}
\begin{subfigure}{0.33\textwidth}
     \centering
    \includegraphics[scale = 0.3]{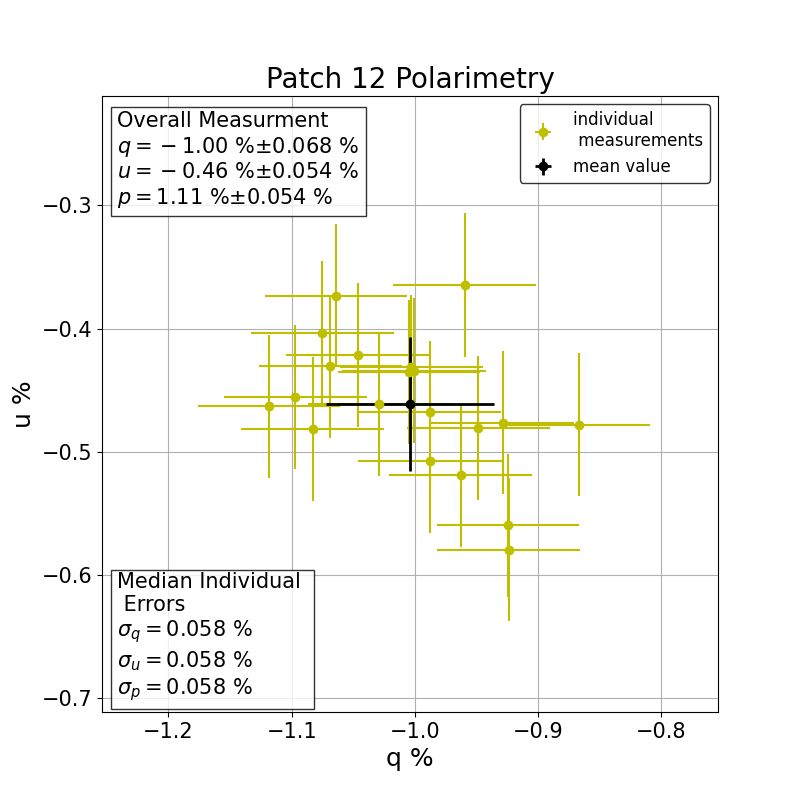}
    \caption{Patch 12}
    \label{qu12}
\end{subfigure}
\begin{subfigure}{0.33\textwidth}
     \centering
    \includegraphics[scale = 0.3]{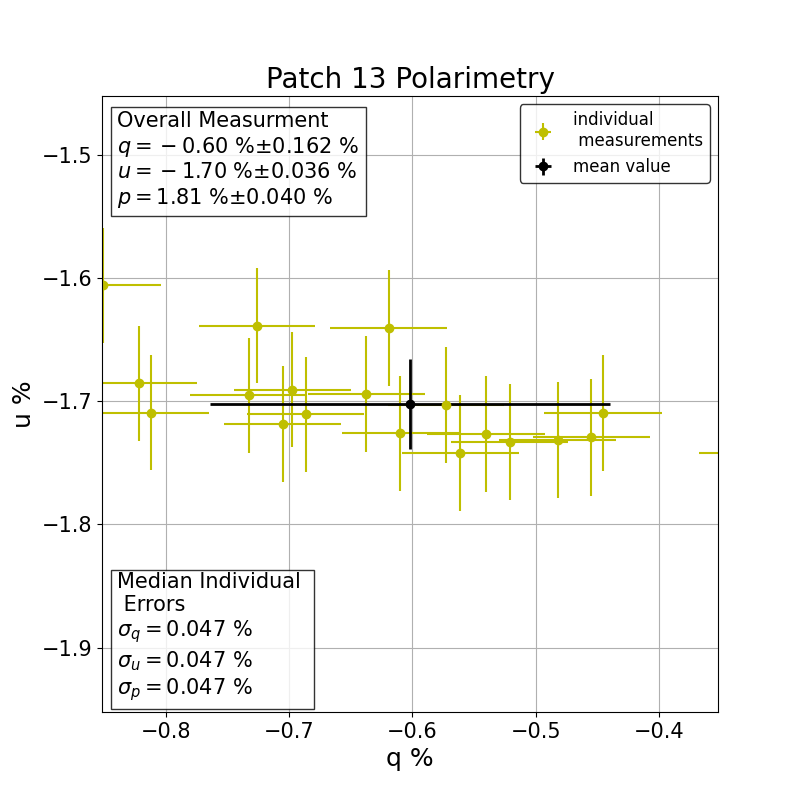}
    \caption{Patch 13}
    \label{qu13}
\end{subfigure}
\begin{subfigure}{0.33\textwidth}
     \centering
    \includegraphics[scale = 0.3]{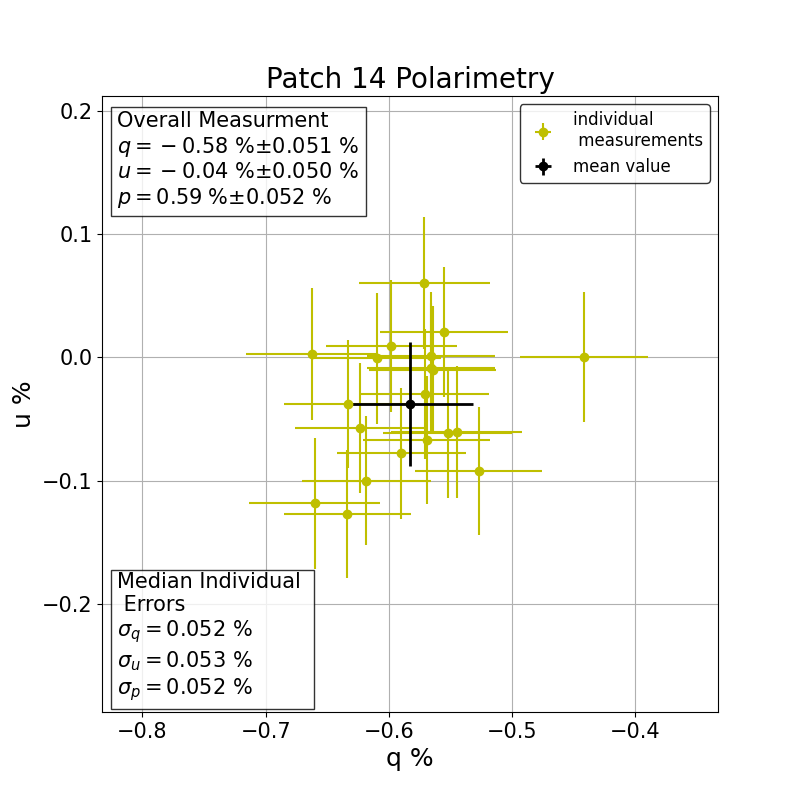}
    \caption{Patch 14}
    \label{qu14}
\end{subfigure}
\begin{subfigure}{0.33\textwidth}
     \centering
    \includegraphics[scale = 0.3]{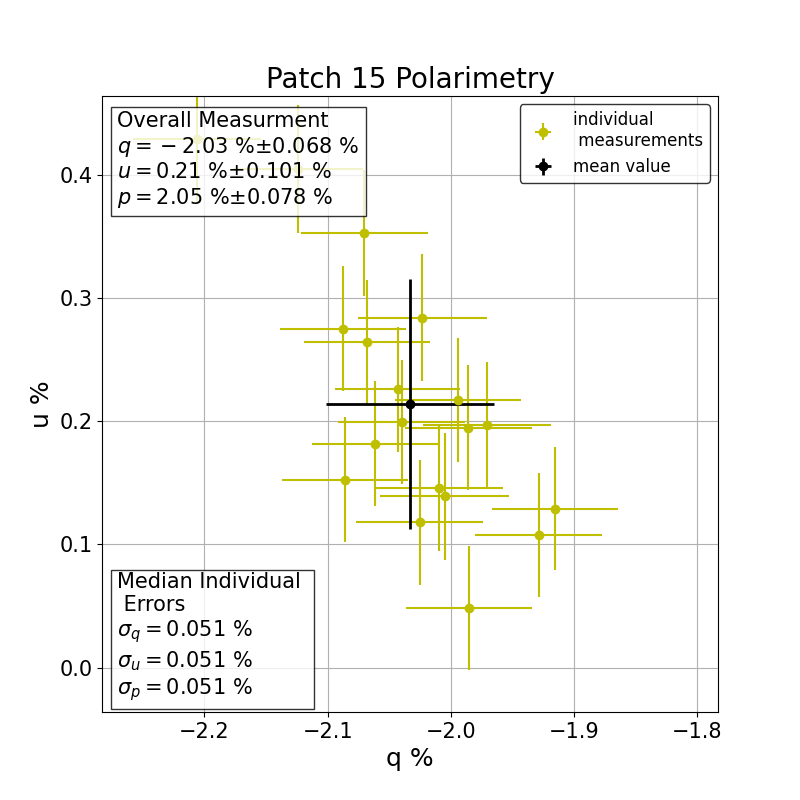}
    \caption{Patch 15}
    \label{qu15}
\end{subfigure}
\begin{subfigure}{0.33\textwidth}
     \centering
    \includegraphics[scale = 0.3]{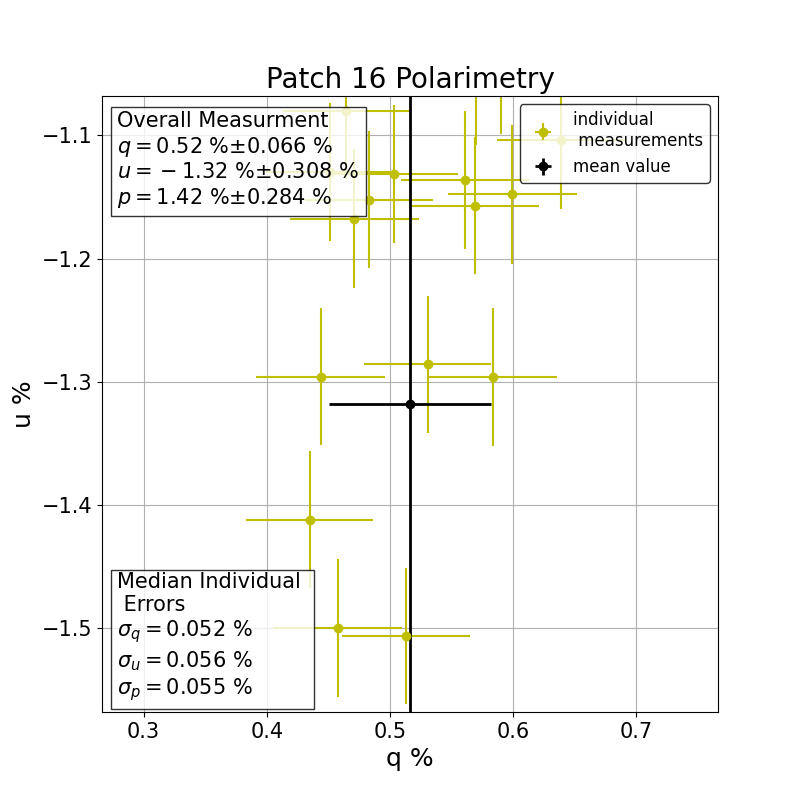}
    \caption{Patch 16}
    \label{qu16}
\end{subfigure}
\begin{subfigure}{0.33\textwidth}
     \centering
    \includegraphics[scale = 0.3]{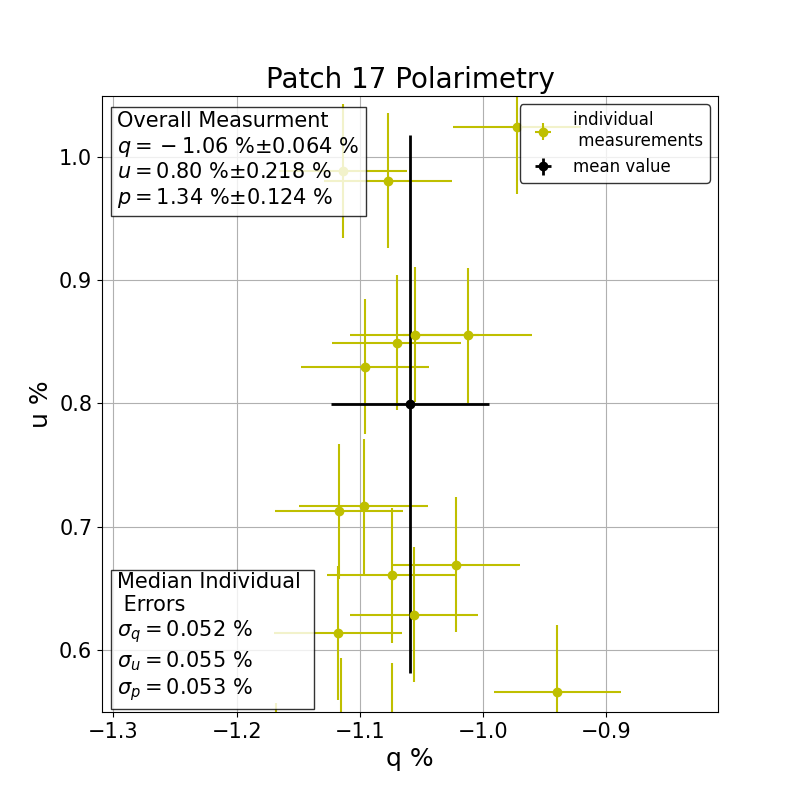}
    \caption{Patch 17}
    \label{qu17}
\end{subfigure}
\begin{subfigure}{0.33\textwidth}
     \centering
    \includegraphics[scale = 0.3]{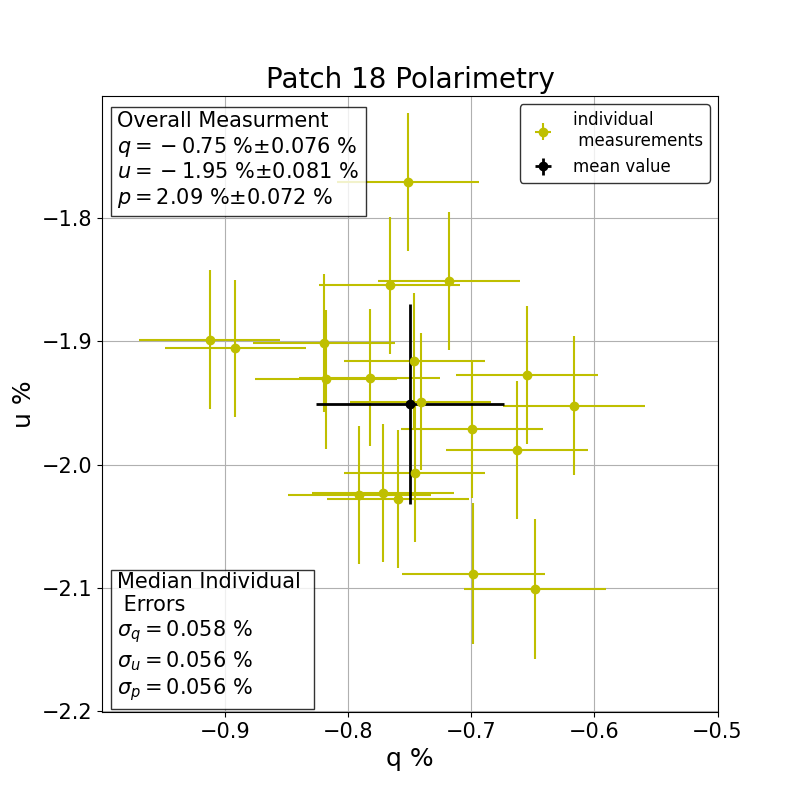}
    \caption{Patch 18}
    \label{qu18}
\end{subfigure}
\begin{subfigure}{0.33\textwidth}
     \centering
    \includegraphics[scale = 0.3]{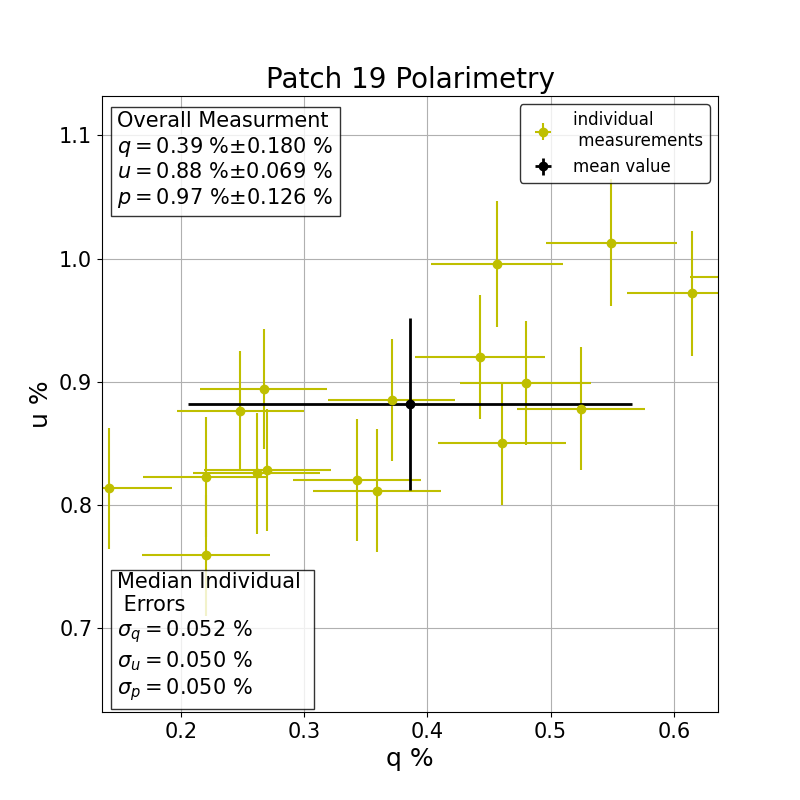}
    \caption{Patch 19}
    \label{qu19}
\end{subfigure}
\caption{Polarimetric measurements for Patches 11 to 19, similar to Fig.~\ref{qu_maps}. Yellow crosses are $q$ and $u$ values for all the individual pointings for a  patch. The overall mean value and the measured standard deviation (black cross) are mentioned in the top left legend.}
\label{fig_app_2}
\end{figure}

Figures~\ref{fig_app_1} and~\ref{fig_app_2} show the measured polarizations and flatness in the $q-u$ plane for Patch~2 to 19.

%
%

\end{document}